\documentclass[usenatbib]{mnras}

\pdfoutput=1
\usepackage{amssymb,amsmath}
\usepackage{graphicx}
\usepackage{multirow}
\usepackage{natbib}
\usepackage{tabularx}
\usepackage[bottom]{footmisc}
\usepackage{wasysym}
\usepackage{upgreek}
\usepackage{color}

\def\arcsec{\hbox{$^{\prime\prime}$}}
\def\arcmin{\hbox{$^{\prime}$}}

\title{The first interferometric detections of Fast Radio Bursts}

\author[M.~Caleb et al.]
{M.~Caleb$^{1,2,3}$\thanks{Email: manisha.caleb@anu.edu.au},
C.~Flynn$^{2,3}$,
M.~Bailes$^{2,3}$,
E.\,D.~Barr$^{2,3,4}$,
T.~Bateman$^{6}$,
S.~Bhandari$^{2,3}$,
\newauthor D.~Campbell-Wilson$^{6,3}$
W.~Farah$^{2}$,
A.\,J.~Green$^{6,3}$,
R.\,W.~Hunstead$^{6}$
A.~Jameson$^{2,3}$,
\newauthor F.~Jankowski$^{2,3}$,
E.\,F.~Keane$^{5}$, 
A.\,~Parthasarathy$^{2,3}$,
V.~Ravi$^{2,3,7}$,
P.\,A.~Rosado$^{2,8}$,
\newauthor W. van Straten$^{2,9}$,
V.~Venkatraman Krishnan$^{2,3}$
\\ \\
$^{1}$ Research School of Astronomy and Astrophysics, Australian National University, ACT, 2611, Australia\\
$^{2}$ Centre for Astrophysics and Supercomputing, Swinburne University of Technology, P.O. Box 218, Hawthorn, VIC 3122, Australia \\
$^{3}$ ARC Centre of Excellence for All-sky Astrophysics\\
$^{4}$ Max-Planck-Institut f{\"u}r Radioastronomie, Auf dem H{\"u}gel 69, D-53121 Bonn, Germany\\
$^{5}$ SKA Organisation, Jodrell Bank Observatory, Cheshire, SK11 9DL, UK\\
$^{6}$ Sydney Institute for Astronomy (SIfA), School of Physics, The University of Sydney, NSW 2006, Australia \\
$^{7}$ Cahill Center for Astronomy and Astrophysics, MC249-17, California Institute of Technology, Pasadena, CA 91125, USA \\
$^{8}$ Monash Centre for Astrophysics, School of Physics and Astronomy, Monash University, VIC 3800, Australia \\ 
$^{9}$ Institute for Radio Astronomy \& Space Research, Auckland University of Technology, Private Bag 92006, Auckland 1142, New Zealand}

\begin{document}
\maketitle

\begin{abstract}
We present the first interferometric detections of Fast Radio Bursts
(FRBs), an enigmatic new class of astrophysical transient. 
In a 180-day survey of the Southern sky we discovered 3 FRBs at 843 MHz 
with the UTMOST array, as part of
commissioning science during a major ongoing upgrade. The wide field
of view of UTMOST ($\approx 9$ deg$^{2}$) is well suited to FRB
searches. The primary beam is covered by 352 partially overlapping fan-beams,
each of which is searched for FRBs in real time with pulse widths in the range
0.655 to 42 ms, and dispersion measures $\leq$2000 pc cm$^{-3}$.
Detections of FRBs with the UTMOST array places a lower limit on their
distances of $\approx 10^4$ km (limit of the telescope near-field)
supporting the case for an astronomical origin. Repeating
FRBs at UTMOST or an FRB detected simultaneously with the Parkes
radio telescope and UTMOST, would allow a few arcsec localisation, thereby
providing an excellent means of identifying FRB host galaxies, if present.
Up to 100 hours of follow-up for each FRB has been carried out with the
UTMOST, with no repeating bursts seen. 
From the
detected position, we present 3$\sigma$ error
ellipses of 15$\arcsec \times 8.4^{\circ}$ on the sky for the point of
origin for the FRBs. We estimate an all-sky FRB rate at 843 MHz above
a fluence $\cal F_\mathrm{lim}$ of 11 Jy ms of $\sim 78$ events
sky$^{-1}$ d$^{-1}$ at the 95 percent confidence level. The measured rate of
FRBs at 843 MHz is of order two times higher than we had expected,
scaling from the FRB rate at the Parkes radio telescope, assuming that FRBs have a flat spectral index 
and a uniform distribution in Euclidean space \citep{Caleb}. We examine how this can be explained by FRBs 
having a steeper spectral index and/or a flatter log$N$-log$\mathcal{F}$ distribution 
than expected for a Euclidean Universe.
\end{abstract}

\begin{keywords}
instrumentation: interferometers -- intergalactic
medium -- surveys -- radio continuum -- methods: data analysis
\end{keywords}

\section{Introduction}

Fast radio bursts (FRBs) are a relatively new class of radio transient
that are short, bright and highly dispersed. The pulses are typically
of durations of a few milliseconds, and exhibit dispersion sweeps
characteristic of propagation through a cold diffuse plasma
\citep{Lorimer,Thornton}. The dispersion measures (DMs) of these
pulses are significantly higher than the contribution from the line-of-sight
through the Galactic Interstellar Medium (ISM), suggestive of a
cosmological origin in which the large DMs are due to passage through
the Intergalactic Medium (IGM). If they are at cosmological distances,
their inferred intrinsic energies ($> 10^{31}$ J) and brightness
temperatures (${T}_\mathrm{b} >10^{33}$ K) necessitate a coherent
emission mechanism, while the short durations of the pulses suggest a
very compact source of origin \citep{Luan, Dennison}.
The 18 FRBs published to date (refer to the FRBCAT 
repository$^{\ref{frbcat}}$ for the complete list) have been discovered in either
post-processing of archival surveys or, in real-time, using
the Parkes radio telescope with
the exception of two, detected at the Arecibo \citep{Spitler} and
Green Bank telescopes \citep[GBT;][] {Masui} All but one of the bursts have
been found at 1.4 GHz, with the exception being the GBT burst, which 
was seen at 800 MHz.

The observed FRB all-sky rate is very high. \cite{Champion} derive a rate of
$7^{+5}_{-3} \times 10^{3}$ events sky$^{-1}$ d$^{-1}$ at 1.4 GHz for bursts
between 0.13 and 1.5 Jy ms in fluence and widths in the
range 0.128 ms to 16 ms. 
The high FRB rate is a major constraint on theories for their origin.
Until recently, such theories have generally assumed they are
cataclysmic events, in which the progenitor is obliterated. However, one
FRB is now known to repeat in a non-periodic manner 
\citep[FRB 121102,][]{nat_spitler}, opening up possibilities for other
progenitor models. 
Following the discoveries reported in this paper, \cite{Chatterjee} have achieved sub-arsecond localisation
of the FRB 121102 using radio interferometric observations from the Very Large Array.
The source has been localised to a $m_{r^\prime} = 25.1$ AB mag low-metallicity, star-forming
dwarf galaxy at $z = 0.19273(8)$ \citep{SriHarsh}. The precise localisation shows
that the source is either co-located with a 180 $\upmu$Jy active galactic nucleus or an unresolved
type of extragalactic source. However, the exact nature of the FRB progenitor is still
unknown.

Despite
concerted follow-up efforts for almost all FRBs,
this remains the only FRB seen to repeat. These efforts have been quite
substantial. For instance, $\approx80$ hrs of followup for the Lorimer burst
\citep{Lorimer}, $\approx80$ hrs for FRB 131104 \citep{Ravi} and $\approx110$ hrs of
selected FRB positions \citep{Petroff_followup} at the Parkes radio
telescope yielded no repeats. This suggests the possibility of there
being two independent classes of FRBs -- repeating and
non-repeating -- with two classes of possible progenitors
\citep{nat_keane}.
Progenitor theories include
flaring magnetars \citep{Lyubarsky}, giant pulses from pulsars
\citep{Wasserman, Connor}, binary white dwarf mergers
\citep{Kashiyama}, neutron star mergers \citep{Totani} and collapsing
supramassive neutron stars \citep{Falcke}. 
It is possible that the lack of repetition of pulses for the FRB discoveries at the Parkes
radio telescope is merely due to limited sensitivity and follow-up time,
and that all FRBs have a common origin \citep{Scholz}.
FRB 010724 is an exception to this however: its extreme brightness ($\sim 30$ Jy) 
far outweighs the lower gain of Parkes relative to Arecibo, so that one cannot infer its 
lack of repeat bursts is due to limited sensitivity. Recently, \cite{RaviSci} 
have reported the detection of FRB 150708, which is of comparable brightness ($\sim 12$ Jy) 
to FRB 010724, and exhibits 100 percent polarisation and suggests weak turbulence 
in the ionised IGM. \cite{DeLaunay} have 
associated a $\gamma$-ray transient with the FRB 131104 discovered by \citep{Ravi}.
However \citep{Ravi2016} in contrast, report on the discovery of a variable source (consistent with an AGN) 
temporally and spatially coincident with the FRB 131104 but not spatially coincident with
the $\gamma$-ray burst, and rule out the association of the $\gamma$-ray burst with the 
FRB using probabilistic reasoning.

\begin{figure}
\centering
\includegraphics[width=0.5\textwidth]{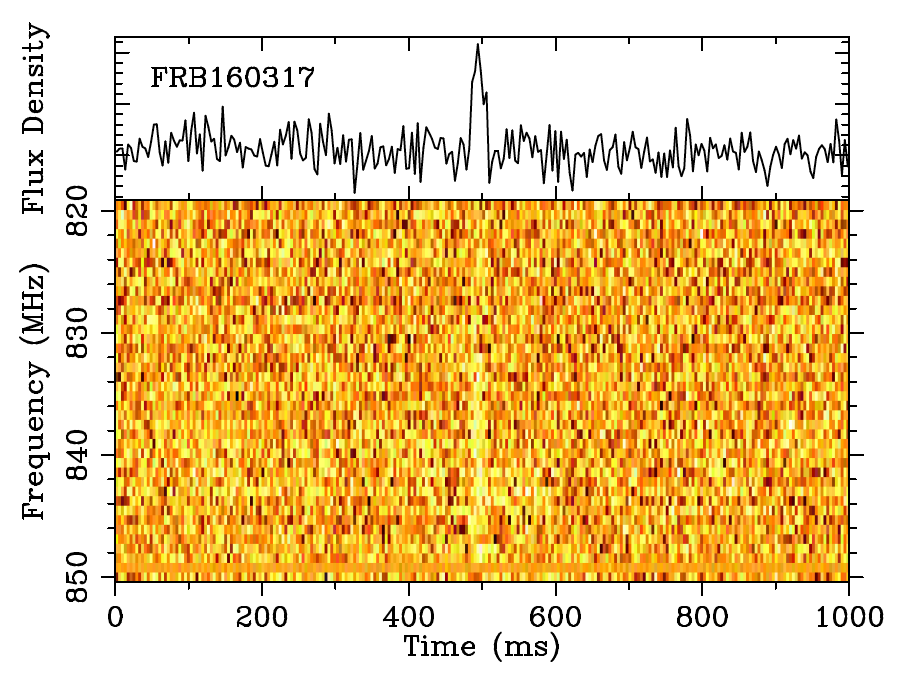}%
\par\medskip
\includegraphics[width=0.5\textwidth]{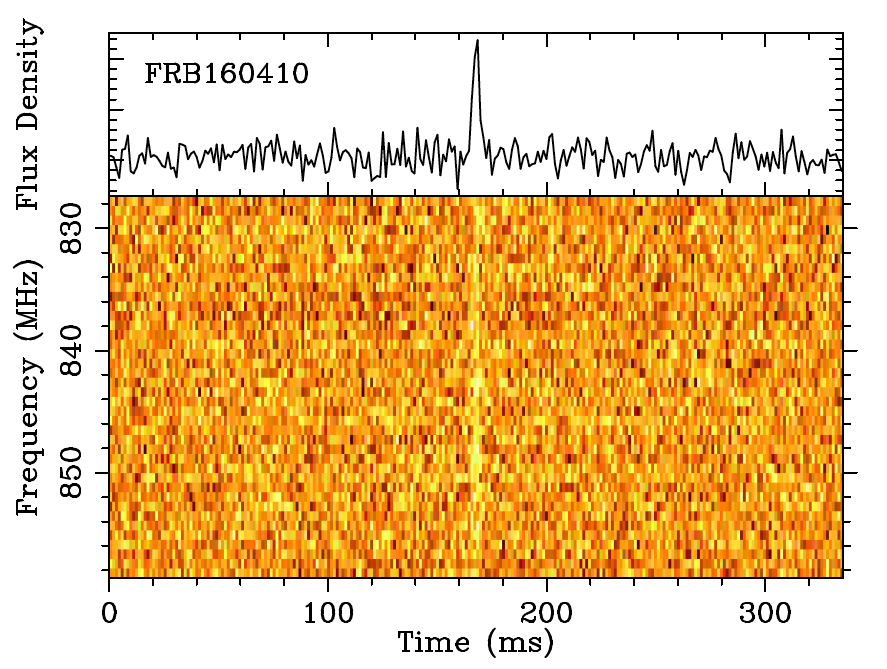}%
\par\medskip        
\includegraphics[width=0.5\textwidth]{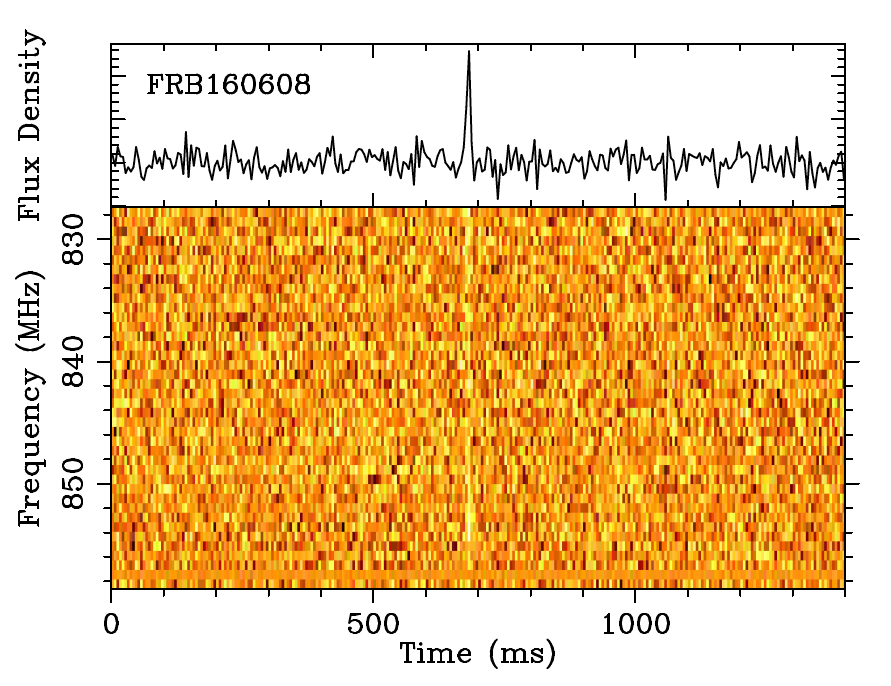}%
\caption{Frequency vs time behaviour of FRBs 160317, 160410 and 160608
  detected at UTMOST at the centre frequency of 834.765 MHz. The top panel in each case shows the frequency-averaged 
  pulse profile. The bottom panel shows that narrow-band RFI has been excised and
  the effects of inter-channel dispersion have been removed assuming
  DMs of $1165 \pm11$, $278\pm3$ and $682\pm7$ pc cm$^{-3}$
  respectively. The data are uncalibrated as the bandpass of the system 
  varies as a function of meridian angle, and the flux densities are in arbitrary units.
  Note the different time range on the abscissa for FRB 160410.}
\label{fig:frbs1}
\end{figure}

Most published FRBs have been detected with single dish antennas, with
relatively poor angular resolution, and we are unable to indisputably
rule out a near-field or atmospheric origin for the one-off events until now. 
The FRB detections made with the multi-beam receiver at the Parkes radio telescope however,
are likely to originate at $\gtrsim 20$ km \citep{Vedantham}.
Also FRB 150418 has been proposed to be associated with a galaxy at
$z \sim 0.5$. However this association has been called into question by \cite{Williams} and
\cite{Vedantham}, and other models like giant pulses from
extragalactic pulsars which could account for the
excess DM in the local environment, have been proposed \citep{Connor}. Better localisation during discovery in the
radio requires an interferometric detection. 

\begin{figure}
\centering
\includegraphics[width=0.5\textwidth]{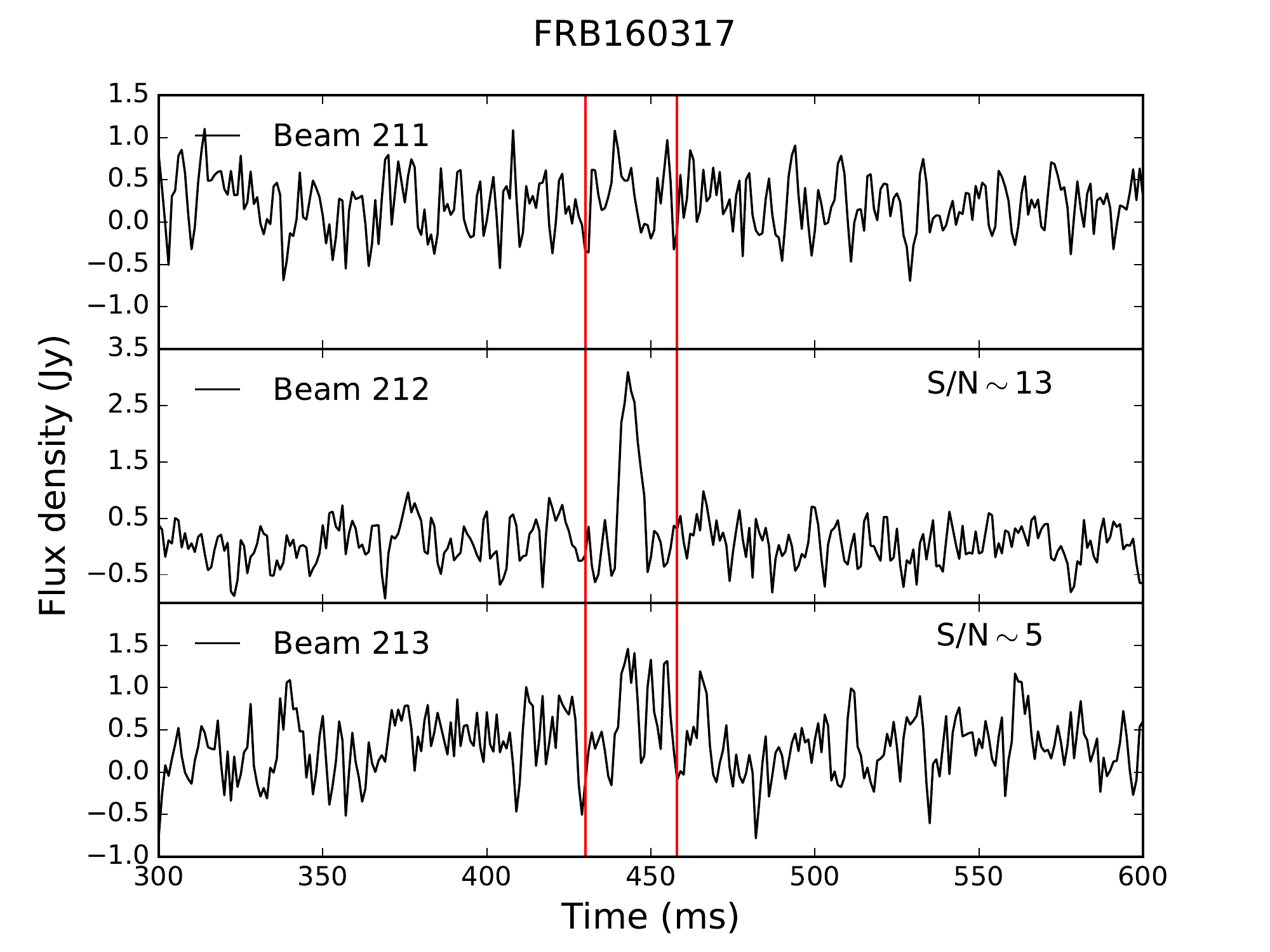}%
\par\medskip
\includegraphics[width=0.5\textwidth]{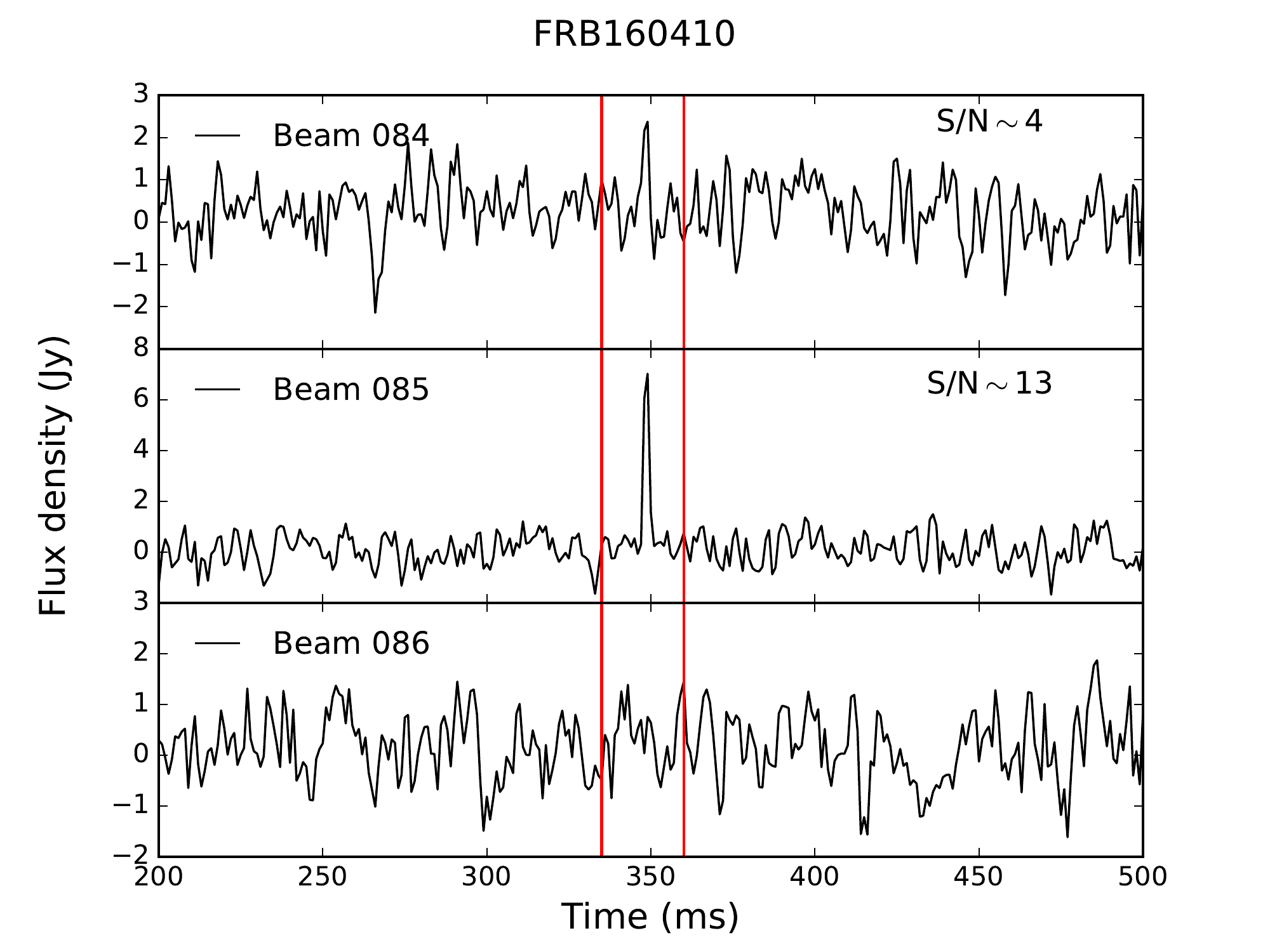}%
\par\medskip        
\includegraphics[width=0.5\textwidth]{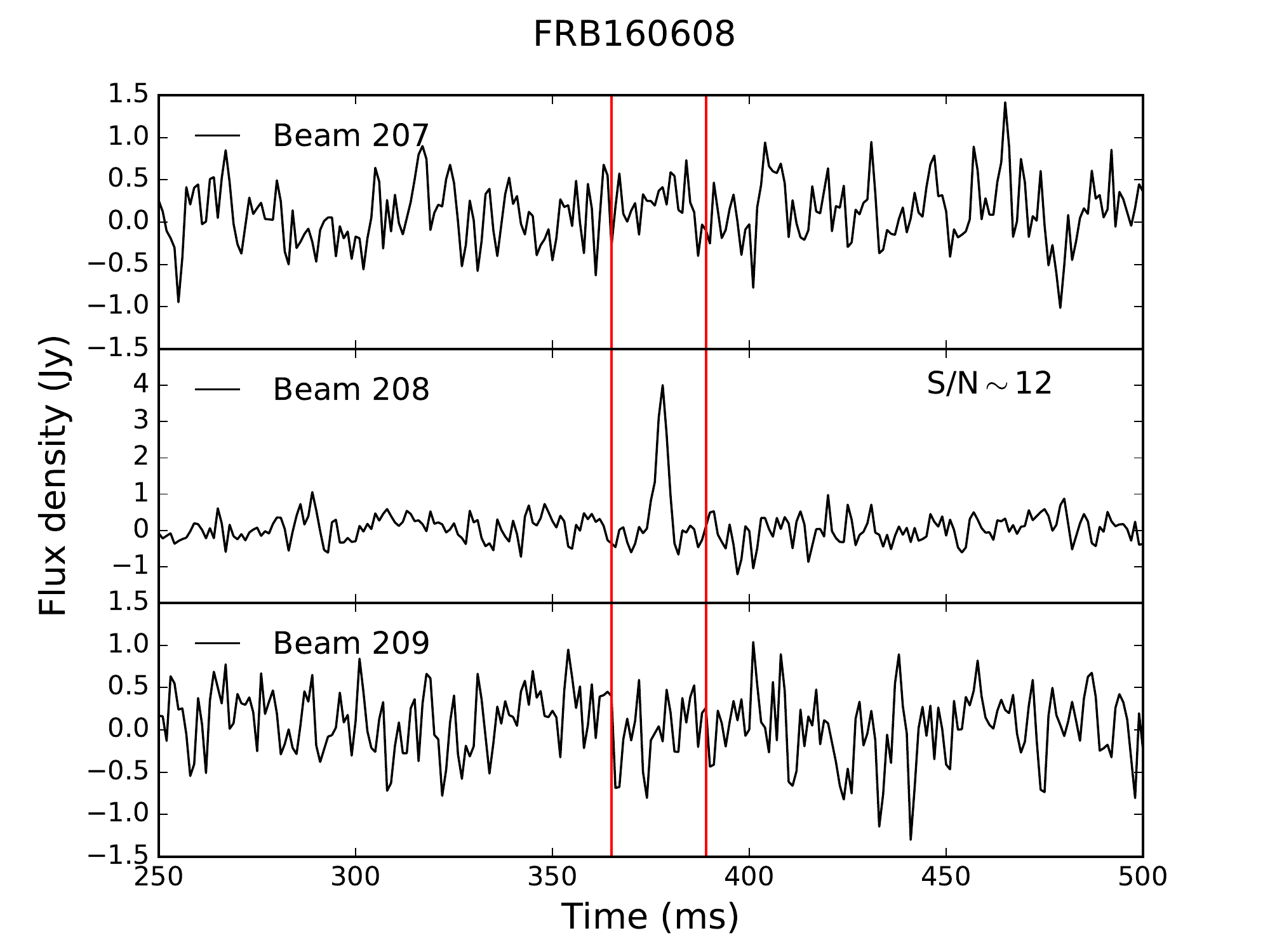}%
\caption{The three panels display the total power pulse profiles for one polarisation
in three adjacent fan-beams. 
FRBs 160317 and 160410 were also detected as sub-threshold events
in neighbouring fan-beams (in addition to the high S/Ns in the primary
detection fan-beams), indicating that they did not occur near the
centres of the primary fan-beam. On the contrary,
FRB 160608 was only detected in one fan-beam suggesting that it
occurred close to the the centre of beam 208 (see bottom panel).}
\label{fig:frbs2}
\end{figure} 

In a companion paper, we describe how the Molonglo Observatory
Synthesis Telescope (sited near Canberra in Australia) is currently
undergoing a major upgrade, with the addition of
a state-of-the-art correlator to transform it into an FRB finding
machine - the UTMOST (Bailes et al., submitted).  Two FRB searches were
performed with UTMOST in 2015 during the upgrade, when the system was
operating at a small fraction of the final expected sensitivity, and
only yielded an upper limit of the FRB rate \citep{Caleb}. 

We have now undertaken a third FRB survey at UTMOST and discovered 3
FRBs. These are the first FRBs observed with an interferometer,
further strengthening the case for an astronomical origin in addition to the 
detections at other telescopes and in the expected number of beams at 
Parkes for far-field events, as detection
with UTMOST implies the events are in the far-field region
$\gtrsim 10^4$ km.  Section \ref{sec:OA}
of this paper briefly outlines the telescope specifications, survey
properties and the transient detection pipeline. We present the bursts'
properties and their follow-up observations and localisation areas in Section
\ref{sec:results}. The event rate estimates of the FRBs at 843 MHz
based on the detections of the 3 FRBs and constraints on their
spectral index are detailed in Section \ref{sec:disc} followed by our
conclusions in Section \ref{sec:conc}.

\section{UTMOST specifications and survey properties}
\label{sec:OA}

The UTMOST consists of an East-West (E-W) aligned cylindrical paraboloid divided into
two `arms' (separated by a 15-m gap), each 11.6-m wide and 778-m long,
with 7744 right circularly polarised ring antennas operating at 843 
MHz on a line feed system at its focus.
Groups of 22 consecutive ring antennas (these groups are termed `modules') 
are phased to the physical centre of the module,
forming 352 unique inputs (each with a beam $4.0^{\circ} \times 2.8^{\circ}$ FWHP) which are then beamformed (Bailes at al., submitted).
We operate the telescope by tilting the arms North-South and steering 
the ring antennae
East-West by differential rotation. UTMOST can access the sky South of $\delta=+18^{\circ}$
with the East-West steering limited to $\pm60^{\circ}$.
The telescope's field of view, sensitivity and high duty cycle make it a near
ideal survey instrument for finding FRBs and other radio transients.
Since late 2015, we have been using UTMOST to search for fast radio
transients for an average of 18 hours a day, while simultaneously
timing more than 300 pulsars weekly (Bailes et al., submitted, 
Jankowski et al., in prep).

In FRB search mode, the 4.0 degrees FWHP of the primary beam is
tiled in the E-W direction by 352 elliptical, coherent,
tied-array beams (called `fan-beams', each 46$\arcsec$ wide), spaced 41$\arcsec$ apart and
overlapping at very close to their half power points at 843 MHz. In the N-S direction
the resolution of the fan-beams is the same as that of the primary
beam ($\approx2.8$ degrees). The fan-beams are numbered from 1 to 352 running
from East to West across the primary beam, with fan-beam 177 directly
centred on boresight.  The sensitivity of the telescope to bursts can
be estimated using the radiometer equation:

\begin{equation}
S_\mathrm{min} = \beta \, \frac {(\mathrm{S/N_\mathrm{min}}) \, T_\mathrm{sys}} {G \, {\sqrt{\Delta\nu \, W \, N_\mathrm{p}}}}
\end{equation}

\begin{figure*}
\includegraphics[width=5.0 in]{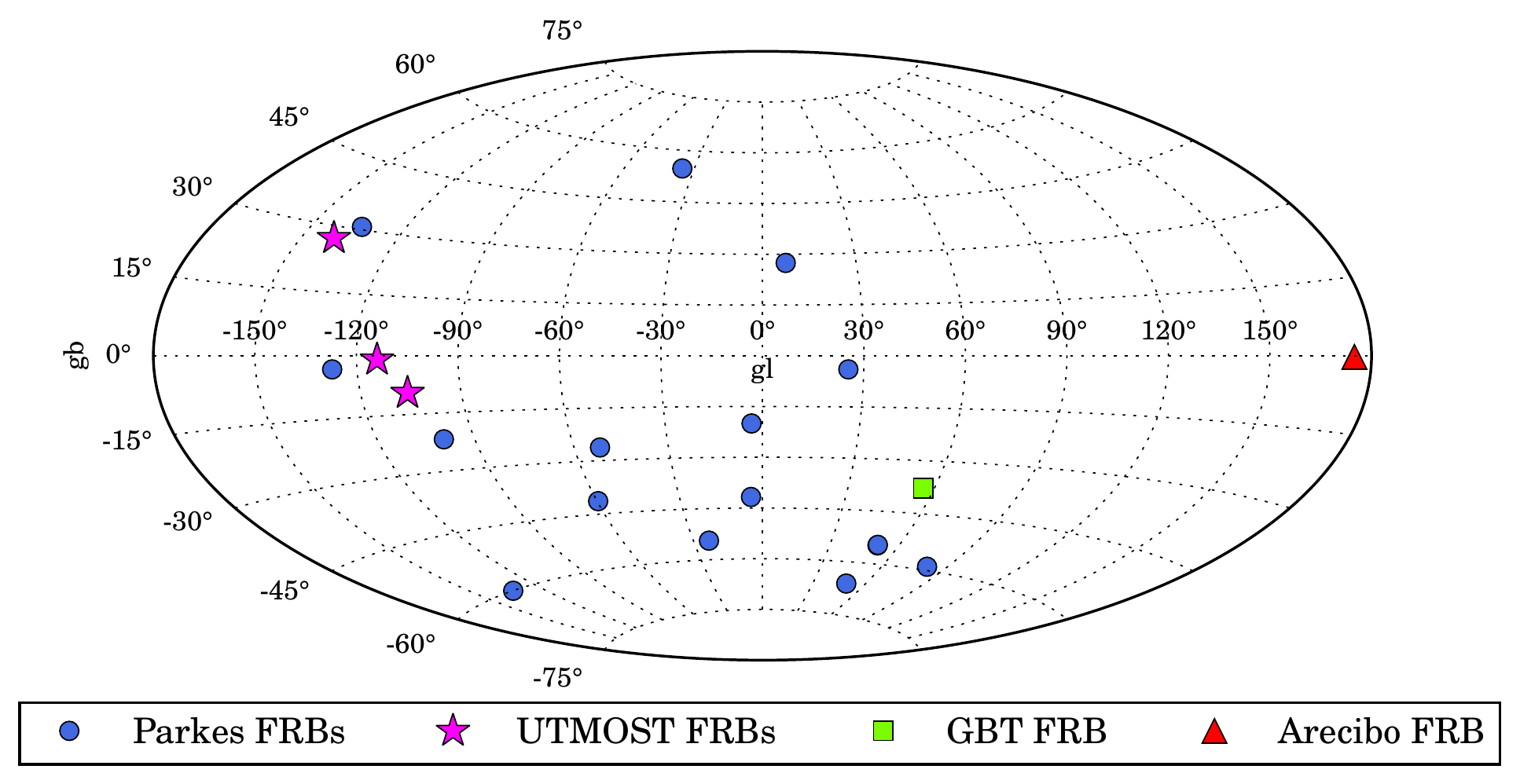}
\caption{The sky distribution of the 18 FRBs published to date in Galactic coordinates. Dots mark the
positions of the FRBs detected at the Parkes telescope, the triangle 
represents FRB 121102 detected at the Arecibo telescope and the square
represents FRB 110523 discovered at the GBT. Stars mark the 
positions of the UTMOST FRBs. Two of the Parkes FRBs have positions separated by 9$\arcmin$ which is not 
resolved in this figure. It should be noted that there are large 
biases in this distribution due to very different sky coverages and survey depths.
}
\label{fig:skydist}
\end{figure*}

\noindent where ${S_\mathrm{min}}$ is the minimum detectable flux for
a threshold signal-to-noise $\mathrm{S/N_{min}}$, $\beta$ is the digitisation factor, $\Delta\nu$
is the bandwidth in Hz, ${N_\mathrm{p}}$ is the number of
polarisations (${N_\mathrm{p}}=1$ for UTMOST as it is right circularly
polarised only), $W$ is the pulse width in ms, ${T_\mathrm{sys}}$
is the system temperatures in K respectively, and $G$ is the system gain in K Jy$^{-1}$.
We define S/N as the ratio of the sum of the on-pulse flux to the product of the
rms of the off-pulse flux and square root of number of on-pulse bins 
($\mathrm{S/N} = \frac{\mathrm{I_{on}}}{\sqrt{\mathrm{nbin}}\,\mathrm{I_{off}}}$).
For the fully upgraded instrument, we expect $S_\mathrm{min} = 1.6$
Jy\,ms for a 10$\sigma$ 1-ms wide pulse, 3.5 K Jy$^{-1}$ gain, 100 K
system temperature and 31.25 MHz bandwidth.
The system bandwidth is however only about half of the initially anticipated 31.25 MHz bandwidth,
as the ring antennas have a significant roll-off in sensitivity away from 843 MHz. 
This has been measured using integrated pulses from the pulsar J1644$-$4559. We find that on average 
$\sim 86$ percent of the total S/N is concentrated in the upper half of the band ($\sim836 - 850$) as the antennas 
are tuned to maximum sensitivity at 843 MHz.
We adopt a bandwidth of 16 MHz for the sensitivity calculations in the paper, to be conservative.

During the upgrade, we characterise the system sensitivity by
a fraction of the final expected gain $\upepsilon$. This factor
encompasses systemic losses due to (1) pointing errors (from
physical misalignment in the modules N-S, and phasing errors in
the antenna system E-W), (2) self-generated radio frequency interference 
(RFI) mainly due to improperly shielded electronics in the receiver boxes near the telescope, 
(3) coherent noise in the receiver boxes,
which affects some sets of adjacent modules, and other inefficiencies
in the system performance that
we are still characterising, such as systematic errors in the
phase/delay solutions across the interferometer (Bailes et al. in submitted).

At present (October 2016), we estimate $\upepsilon \approx 0.14$, based
on observations of strong calibrators of known flux densities and a number of
high DM pulsars with relatively stable flux densities. This implies an
effective $T_\mathrm{sys}$ of $400 \pm 100$ K. This is
significantly higher than the system temperature seen on the best
performing modules, which can be as low as 100 K. We note that
$\upepsilon$ can vary from day to day as modules are either
serviced in the field or have electronics maintenance in the
workshops, and typically lie in the range $0.15 < \upepsilon <
0.20$. Occasionally, if only one arm is operational, we have the option to continue surveys
at half sensitivity (i.e. $0.07 < \upepsilon < 0.10$). 
The telescope can access the Southern sky for $\delta < +18^{\circ}$,
and for most parts of the sky we tend to observe reasonably close to
the meridian, in order to maximise sensitivity. The
sensitivity is reduced by projection effects away from the meridian.

\begin{figure*}
\includegraphics[width=0.47\textwidth]{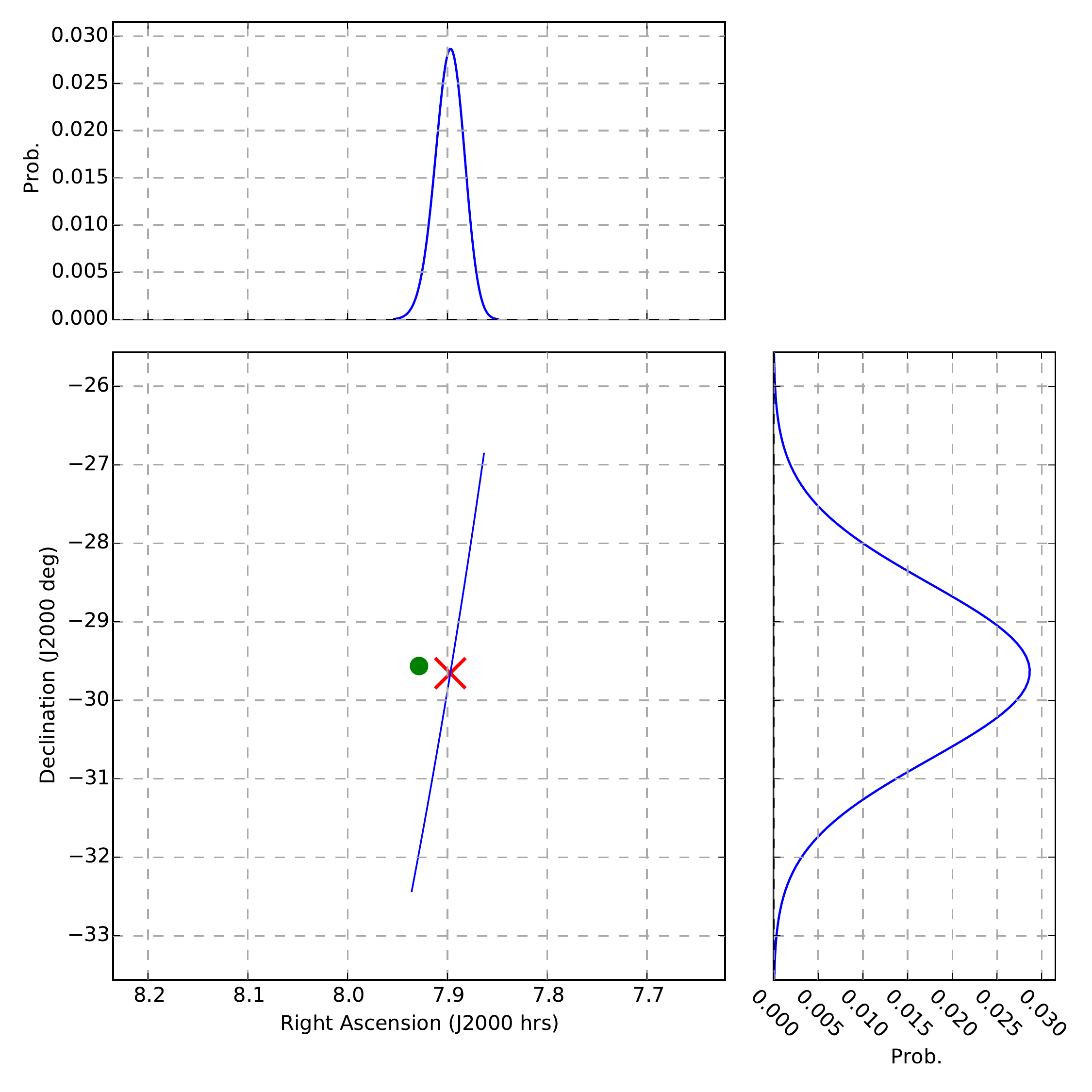}\quad
\includegraphics[width=0.47\textwidth]{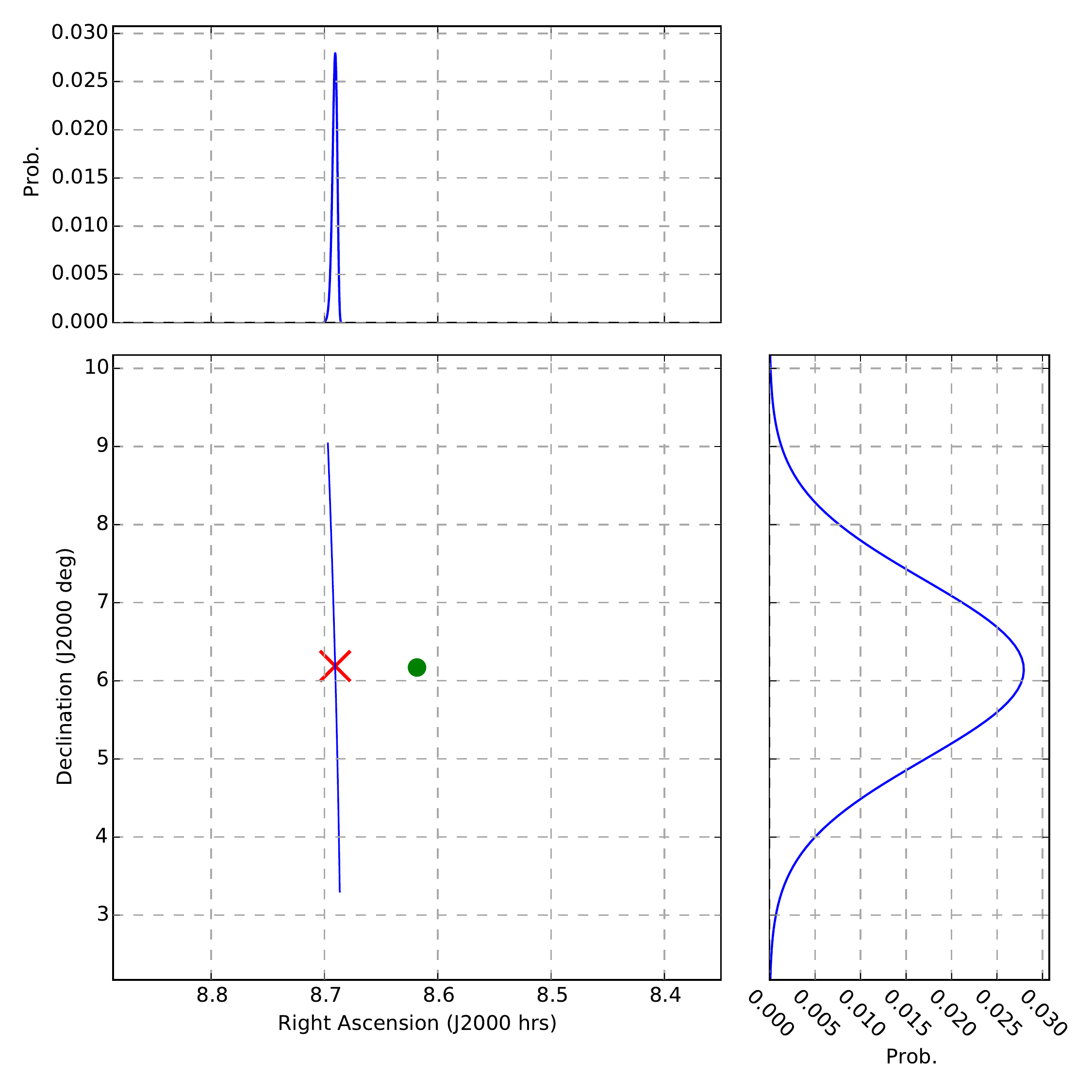}
\medskip
\includegraphics[width=0.47\textwidth]{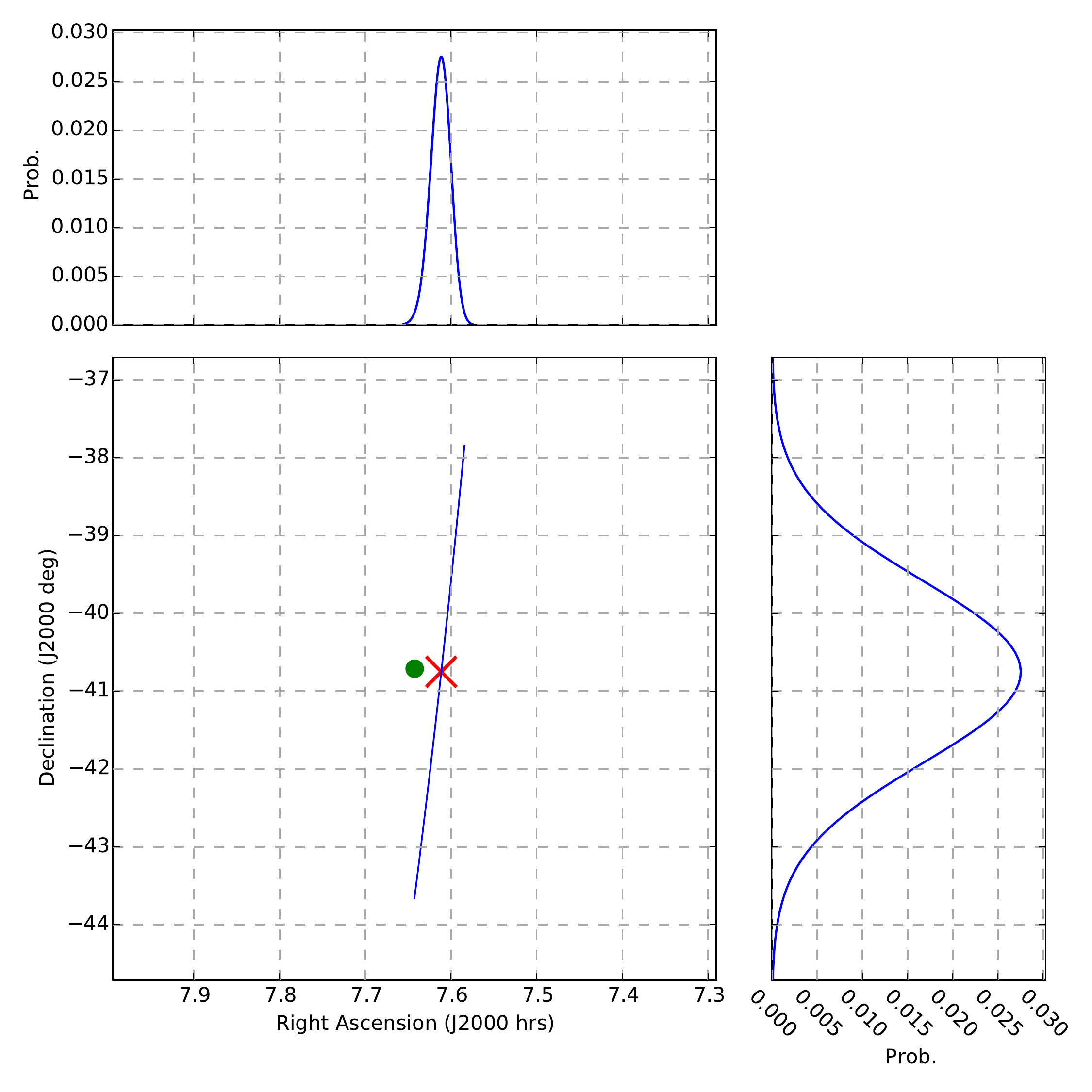}
\caption{We show 3$\sigma$ (15$\arcsec \times 8.4^{\circ}$) localisation ellipses of FRB 160317 (top-left), FRB 160410 (top-right) and FRB 160608 (bottom-centre). The maximum probability in RA (J2000) and DEC (J2000) assuming a Gaussian probability density function gives the most likely position of the FRB, and is marked by the cross. The dot indicates the position of the ``boresight'' pointing of the telescope. Typically, a pulsar is being timed at this position in the telescope beam. In 2 out of 3 cases, the pulsar at this position was bright enough to see individual pulses during the observation when a FRB was detected. The coordinates of the ellipses are given in Table \ref{table:coords}.}
\label{fig:frbspos}
\end{figure*}

\label{sec:pipeline}

In November 2015, we commenced our third FRB survey ``V3.0''. 
It ran for a total of 159.0 days on sky
(between 01-11-2015 and 30-11-2016), at $\upepsilon \approx 0.14$ of the final
target telescope sensitivity. Our fluence limit of the survey,
that is the fluence of the narrowest detectable pulse $\cal{F}_\mathrm{lim}$ can be parametrized as,

\begin{equation}
\label{eq:lim_flue}
{\cal{F}_\mathrm{lim}} \approx 11 \Bigg(\frac{W}{\mathrm{ms}}\Bigg)^{1/2}\,  \mathrm{Jy \, ms}
\end{equation}

\noindent where, 11 Jy is the UTMOST flux limit for $\mathrm{S/N}=10$, $G=3.0$ K Jy$^{-1}$, $\Delta\nu=16$ MHz,
$W=1$ ms, $N_\mathrm{p}=1$ and $T_\mathrm{sys}=400$ K. 
It should be noted that this is not the same as the fluence completeness limit $\cal{F}_\mathrm{complete}$. 
Between $\cal{F}_\mathrm{lim}$ and $\cal{F}_\mathrm{complete}$ we are incomplete and not all FRBs with fluences 
in this range are detectable. This incompleteness region corresponds to the pink shaded region in Figure \ref{fig:fluence}.
The two previous surveys (V1.0 and V2.0) reported in \cite{Caleb}
yielded no FRB events. Relative to V3.0, V1.0 ran for 19.5 days at
lower sensitivity ($\upepsilon = 0.07$), while
V2.0 operated for 9.4 days at the same sensitivity ($\upepsilon =
0.14$).
FRB survey V3.0 consists primarily of pointings taken commensally
during pulsar timing observations. In this mode, the time series data from 352 fan-beams are searched 
for dispersed single pulses in real time, using a custom version of the heimdall software on 8 Nvidia 
GeForce GTX TITAN X (Maxwell) GPUs with a latency of 8-s. The resulting
candidates were then processed offline, typically the following morning
for overnight pulsar timing (RFI is much reduced at night, and the
telescope is made available for maintenance on week days). On
weekends, the telescope is usually operated continuously. The
candidate processing pipeline used is described in detail in
\cite{Caleb}. The process followed is:

\begin{enumerate}

\item obtain 352 data streams (8-bits/sample), one for each fan-beam, at 655.36-$\upmu$s sampling

\item search time series for single pulses with width, $0.65536 < W <
  41.943$ ms ($W = 2^\mathrm{N} \times 0.65536$ ms, where N =
  0,1,2...) and DMs in the range $100 < \mathrm{DM} < 2000$ pc
  cm$^{-3}$,

\item remove events occurring simultaneously in more than 3 fan-beams
  at a given instant in time, 

\item classify only events with S/N $\geq10$, DM $\geq100$ pc cm$^{-3}$
  and $W\leq41.943$ ms as potential FRB candidates. These then require
  human scrutiny of the diagnostic plots, to remove candidates that were
  RFI, almost always due to narrow-band mobile handset
  emissions in our operating passband and single pulses from known
  pulsars.

\end{enumerate}

\section{Results}
\label{sec:results}
 
The false positive rate at UTMOST is high due to RFI caused by mobile
phone handsets, which produce narrow band (5-MHz) emission in our
band, typically in $\approx 20$ ms pulses. These can be eliminated because
celestial pulses are expected to be broadband,
modulated by a frequency dependent response across the 31.25 MHz bandwidth.
This process has been validated using individual pulses from about 20 bright
pulsars seen to date. We are presently automating this process using machine
learning algorithms, so that pulses can trigger a full voltage dump of the raw
data while they are still in the $\approx30$ seconds of ring
buffer storage, with alerts issued in near real-time. RFI occurs
predominantly at low DM, but the rate is high enough to produce a few
hundred spurious candidates above our DM limit of 100 pc\,cm$^{-3}$
daily. Candidates were typically vetted each morning after data
taking.

In 2016 March, April and June we made the first interferometric
detections of FRBs at 843 MHz: FRB 160317, FRB 160410 and FRB 160608,
as shown in Figure \ref{fig:frbs1}.

\subsection{FRB 160317}
This was detected on 2016 March 17 at 09:00:36.530 UTC
  while observing an X-ray magnetar SGR 0755$-$23, in response to
  an Astronomers Telegram \citep{Barthelmy}. The burst occurred about 0.4 degrees East 
  of the magnetar, and was
  detected $\sim$1$^{\circ}$ off the Galactic plane with a DM of 1165(11)
  pc cm$^{-3}$. The DM due to the ISM at this sight-line is
  $\sim320$ pc cm$^{-3}$ from the NE2001 model by
  \cite{Cordes} and $\sim395$ pc cm$^{-3}$ from the YMW16 model \citep{Yao}. 
  The burst with S/N $ \sim 13$, occurred East of
  the centre of the primary fan-beam of detection (Beam 212) since it appeared weakly 
  in the adjacent fan-beam with S/N $\sim 5$ (Beam 213) as shown in Figure \ref{fig:frbs2}. 

\subsection{FRB 160410}
 Similarly to FRB 160317, this FRB was also
  detected in two adjacent fan-beams (Beam 085 with S/N
  $\sim 13$ and Beam 084 with S/N $\sim 4$) as seen in Figure
  \ref{fig:frbs2}. A single dispersed pulse was discovered on 2016
  April 04 at 08:33:39.680 UTC, in an observation of the pulsar
  J0837$+$0410 at the telescope's boresight. This pulsar is
  so bright that individual pulses were seen from it
  as the FRB occurred, meaning the flux density scale
  and bandpass response of the observation were well understood. The
  FRB was seen $\sim 1^{\circ}$ away from boresight. This pulse was
  detected at Galactic latitude, $\sim27^{\circ}$ 
  with the line-of-sight DM accounting for only $\sim58$ pc cm$^{-3}$
  of the total observed DM from the NE2001 model. The YMW16 model 
  estimates $\sim 63$ pc cm$^{-3}$. FRB 160410 has the lowest DM excess
  $\sim220$ pc cm$^{-3}$ of any published FRB, potentially
  making it the closest FRB discovered to date 
  and an excellent candidate to search for repeat pulses.

\subsection{FRB 160608} The burst occurred in an
  observation of the pulsar J0738$-$4042 at $l = 254.11$ deg and
  $b=-9.54$ deg on 2016 June 06 at 03:53:01.088
  UT with a total DM of $\sim682$ pc cm$^{-3}$ and $\sim238$ pc
  cm$^{-3}$ contribution from the Milky Way (NE2001). The YMW16 model's 
  estimate however is $\sim 310$ pc cm$^{-3}$.
  It was seen $\sim0.5^{\circ}$ 
  from the boresight position. FRB 160608 was detected with
  S/N $\sim 12$, just above the detection threshold of 10 and it
  occurred towards the centre of the primary detection fan-beam
  (Beam 208). No pulse was detected in the
  adjacent fan-beams (see Figure \ref{fig:frbs2}). This
  was initially of concern, but tests with the Vela pulsar placed sufficiently
  far South of the telescope boresight, to produce an individual pulse
  with the same S/N showed that detection in a single
  fan-beam occurred $\approx20\%$ of the time. The localisation of this FRB is thus slightly poorer ($21\arcsec \times 8.4^{\circ}$) than for the other two FRBs, for which 2 fan-beam detections allow more accurate positions. \\

\begin{table*} 
\centering
\label{tab:specs}
\caption{Table of observed and inferred properties of the 3 FRBs in this work. The UTCs are the start times of the observations and the times at which the events occurred. Sky coordinates are the most likely positions of the FRB event within a narrow error ellipse (see Figure \protect\ref{fig:frbspos}). The peak fluxes ($S_\mathrm{peak, obs}$) are computed using the radiometer equation and the DM contribution from the Milky Way (DM$_\mathrm{Gal}$) is calculated using the NE2001 model \citep{Cordes}. The ``boresight fluence'' is the detected fluence corrected for the primary beam and fan-beam responses. They are shown as lower limits, for the unknown correction to higher fluence along the semi-major axis of the detection fan-beam. The isotropic energy $E_{0}$, is the energy at source and $\uptau_{843 \,\mathrm{MHz}}$ is the DM smearing due to the pulse broadening caused by the incoherent dedispersion at the observing frequency. The observed widths and their uncertainties are measured using the \textsc{destroy}$^{\ref{destroy}}$ single pulse search software, \textsc{psrchive}$^{\ref{psrchive}}$ and scripts made publicly available through the FRBCAT repository$^{\ref{frbcat}}$. The redshift $z$ is computed as $(\mathrm{DM}_\mathrm{FRB}-\mathrm{DM}_{\mathrm{Gal},\mathrm{NE2001}})/1200$ \citep{Ioka, Inoue}. The luminosity and co-moving distances are calculated for a standard, flat-universe $\Lambda$CDM cosmology using \textsc{cosmocalc} \citep{Wright}. The boresight sources are the magnetar or pulsars that were being observed during the time the FRB occurred.} 
\begin{tabular}{c c c c}
\hline\hline
Parameter                   & FRB 160317 & FRB 160410 & FRB 160608 \\ [0.5ex] 
\hline
UTC start & 2016-03-17-08:30:58 & 2016-04-10-08:16:54 & 2016-06-08-03:52:24 \\ 
UTC event & 2016-03-17-09:00:36.530 & 2016-04-10-08:33:39.680 & 2016-06-08-03:53:01.088  \\
RA J2000 (hh:mm:ss) & 07:53:47 & 08:41:25 & 07:36:42 \\
DEC J2000 (dd:mm:ss) &  $-$29:36:31 & +06:05:05 & $-$40:47:52 \\
$l$ (deg) & 246.05 & 220.36 & 254.11 \\
$b$ (deg) & $-$0.99 & 27.19 & $-$9.54 \\
Detection S/N & 13 & 13 & 12 \\
$S_\mathrm{peak, obs}$ (Jy) & $>3.0$ & $>7.0$ & $>4.3$ \\
Boresight Fluence (Jy\,ms) &  $>69$ &  $>34$  & $>37$ \\
Isotropic energy, $E_{0}$ (J) & $\sim 10^{34}$ & $\sim 10^{32}$ & $\sim 10^{33}$ \\
Observed width, $W$ (ms)  & 21(7)  & 4(1) & 9(6) \\
DM smearing, $\uptau_{843 \,\mathrm{MHz}}$ (ms) & 12.6 & 3.0 & 7.4 \\
$\mathrm{DM}_\mathrm{FRB}$ (pc cm$^{-3}$) & 1165(11) & 278(3) & 682(7) \\
$\mathrm{DM}_{\mathrm{Gal},\mathrm{NE2001}}$ (pc cm$^{-3}$) &  319.6  & 57.7 & 238.3 \\
$\mathrm{DM}_{\mathrm{Gal},\mathrm{YMW16}}$ (pc cm$^{-3}$) &  394.6  & 62.5 & 310.3 \\
Inferred redshift, $z$ & 0.7 &  0.2 &   0.4 \\
Luminosity Distance (Gpc) & 4.30  & 0.89  & 1.97 \\
Co-moving Distance (Gpc)  & 2.52  & 0.75 & 1.44 \\
Boresight source & SGR 0755$-$2933 & J0837$+$0610 & J0738$-$4042 \\[1ex] 
\hline  
\end{tabular} 
\end{table*}

\begin{figure}
\includegraphics[width=3.5 in]{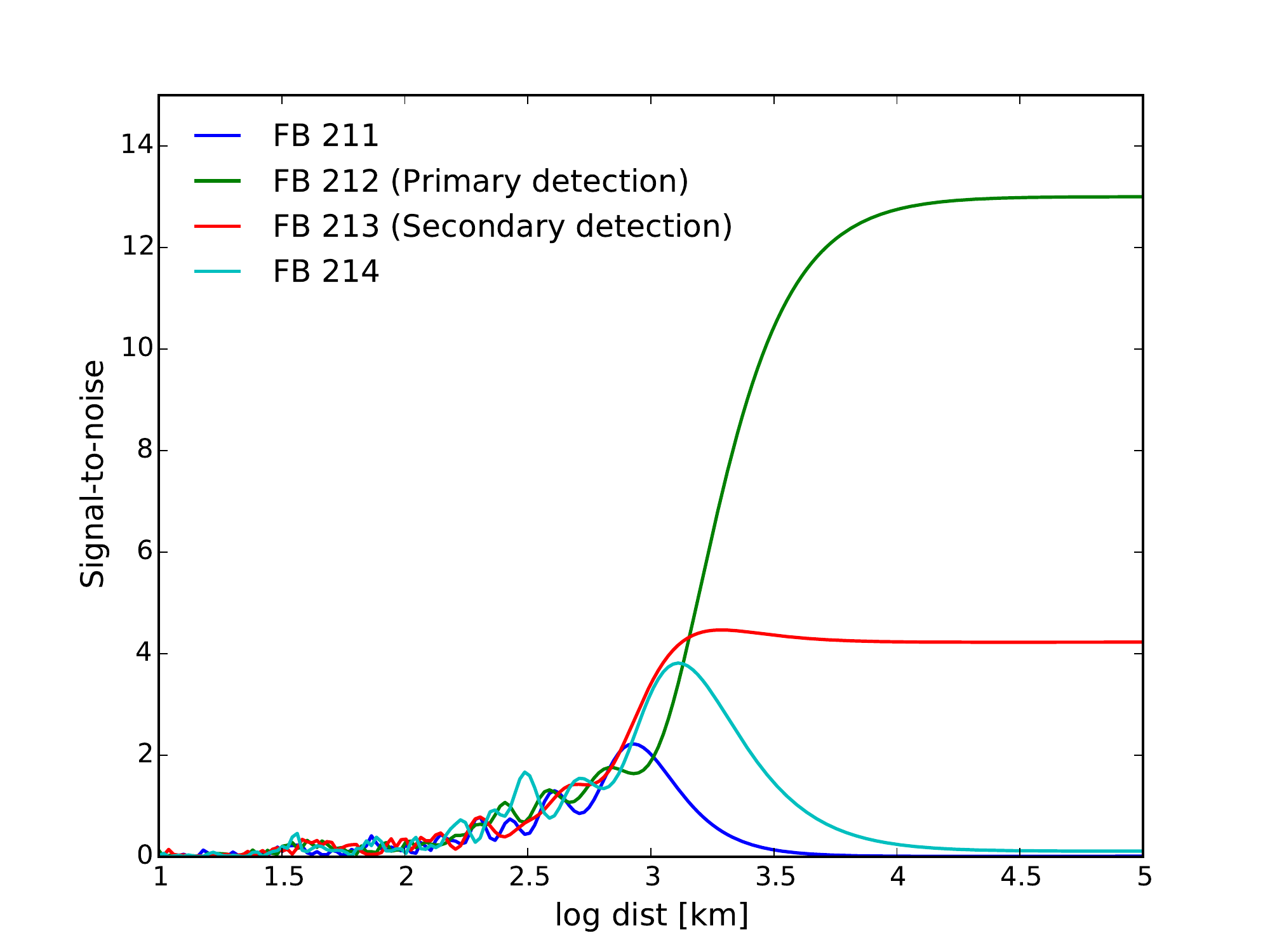}
\caption{Simulations of the detections of FRB 160317 in adjacent fan-beams (FB) to 
determine the Fresnel limit of the telescope. At a distance of $\gtrsim 10^{4}$ km,
the S/Ns of the modeled pulse in FB 212 and FB 213, match that of the observations
with non-detections in the other fan-beams.}
\label{fig:fresnel}
\end{figure}

The primary advantage of the array is that a pulse from a far-field
point source is detected in a maximum of 3 adjacent fan-beams at any
given time, confirmed by extensive pulsar
observations. RFI is typically
near-field, and predominantly appears in more than 3 adjacent
fan-beams, meaning that it can be reliably excised to reduce false
positive rates when searching for transients. Using the 
adjacent fan-beam detections of FRB 160317, we have 
modeled the point of separation between the near-field or Fresnel
region and the far-field or Fraunhofer region of the telescope.
Assuming a point source at $10^{6}$ km, we compute the S/N for a
tied-array beam (e.g. fan-beam 212) phased at an offset of 0.3 from
the centre of the beam to ensure a two fan-beam detection. We compute 
the path length to each module, the phase 
of the signal along the array and perpendicular to the array, 
and add all these as a vector sum weighted by the module performance, to 
get the ``boresight'' S/N. We see that in Figure \ref{fig:fresnel} at a distance of $\gtrsim 10^4$ km, 
we achieve a two fan-beam detection with S/N $\sim 13$ in the primary 
detection beam and S/N $\sim 5$ in the secondary detection beam, similar to the
FRB being modeled. Detections of FRBs in 1 or 2
fan-beams only, thus allows us to identify them as sources more distant
than this, placing them well away from the Earth and hence effectively rule out sources 
of local origin.

The discovery observations containing the FRBs were carefully inspected
to check for similar events at the same time and with the same DM as
the FRB, in other fan-beams. No other broadband pulses
were detected in any other fan-beams within approximately 60 seconds
of the bursts. Moreover, in addition to all the tied array fan-beams, 
we form a single special fan-beam as the incoherent sum of all the
other fan-beams. This ``total power'' fan-beam was also searched for
events near the UTC of the 3 bursts. For the 3 FRBs, this fan-beam
contained no unusual sources of RFI. Only twice
during the 3 surveys did we find FRB-like
candidates (i.e. appearing across the band) which were identified as
RFI upon closer analysis. In each case, similar events could be
found in dozens to hundreds of fan beams, and
were thus obvious near-field RFI. These false candidates also had `patchy'
power across the observing band, indicative of RFI generated from 
different carrier handsets operating at the same time in our band.

\addtocounter{footnote}{1}\footnotetext{\url{https://github.com/evanocathain/destroy\_gutted}
\label{destroy}}
\addtocounter{footnote}{1}\footnotetext{\url{http://psrchive.sourceforge.net/}
\label{psrchive}}
\addtocounter{footnote}{1}\footnotetext{\url{https://github.com/frbcat/FRBCAT\_analysis}
\label{frbcat}}

\begin{figure*}
\includegraphics[width=6.0 in]{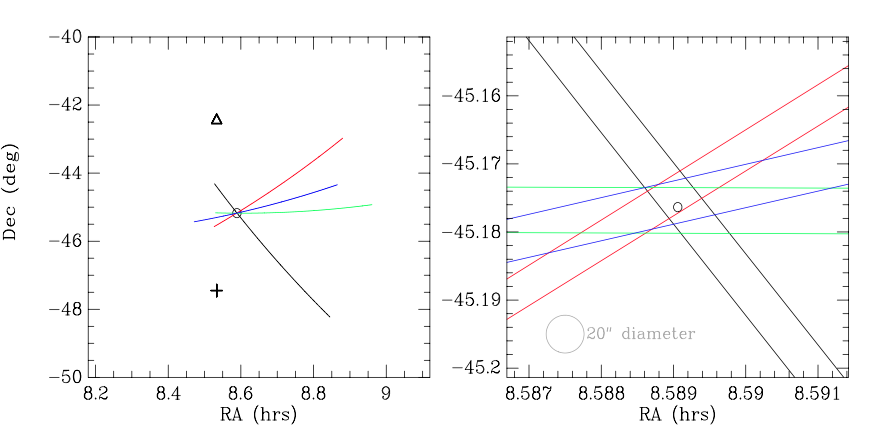}
\caption{Left panel: Localisation contours for four single pulses from Vela,
  observed over different hour angles and distances from the
  telescope boresight. The triangle marks the boresight position for the 
  black fan-beam and the cross marks the boresight position for the three other fan-beams.
  Right panel: A zoom into the
  3$\sigma$ error ellipse for the position of the source on the sky. 
  The circle indicates the position
  of the pulsar. Even a single repeat of an FRB at a different hour angle,
  could constrain the position to a few arcsec
  radius shown in localisation contours in the right panel.}
\label{fig:vela}
\end{figure*}

Two of the three FRBs have been discovered relatively close to the Galactic
plane, with the locations marked as stars in
Figure \ref{fig:skydist}. All three have DMs significantly in excess
of the Galactic contribution, suggesting an extragalactic or
cosmological origin. Under this assumption, the contribution from the
IGM to the DM can be used to infer a redshift,
using the scaling relation in \cite{Ioka} and \cite{Inoue}. This places
FRBs 160317, 160410 and 160608 at a redshift upper limit of 0.7, 0.2 and 0.4
respectively, assuming zero contribution from any potential host galaxy. 
Any contribution from a host galaxy or the
immediate vicinity of an associated source, could be a
significant fraction of the total DM depending on its orientation and
location. The average DM for elliptical galaxies is 37 
pc cm$^{-3}$ and for spiral galaxies is 45 pc cm$^{-3}$ based on the 
probability distribution of DMs computed for a range of host galaxies \citep{Xu}.
For spirals, the weighted average over a range of inclination angles is
estimated to be 142 pc cm$^{-3}$. However the host contribution to the DM
from high redshift galaxies
can be small due to cosmological time dilation and the corresponding 
redshifting of frequency \citep{Zhou}. It also does not account for any bias 
in FRB locations within galaxies.

The S/Ns, DMs and widths of all three FRBs have been computed using
the \textsc{destroy}
single pulse search software,
\textsc{psrchive} with
scripts made publicly available through the FRBCAT
repository. 
The observed widths of all three FRBs 
are dominated by dispersion smearing as shown in 
Table \ref{tab:specs}. This is due to our small bandwidth and 
limited number of channels (40 channels).
We have now implemented a fine channel 
mode (320 channels) which will potentially increase our sensitivity and 
the FRB detection rate by a factor of $\sqrt{8}$. Our total bandwidth of only 31.25 MHz is too
narrow to permit a measurement of dispersion index. 
Single pulses from the Vela pulsar were used to test our
sensitivity to the DM index. The DM 
and the DM index $\delta$ where the dispersive delay is given by,
\begin{equation}
\Delta t \propto \Delta\nu^{-\delta}
\end{equation}
\noindent are found to be highly correlated, so that we can place no practical 
limit on $\delta$. We therefore set the DM index to $\delta = -2$.

\begin{figure*}
\includegraphics[width=5.0 in]{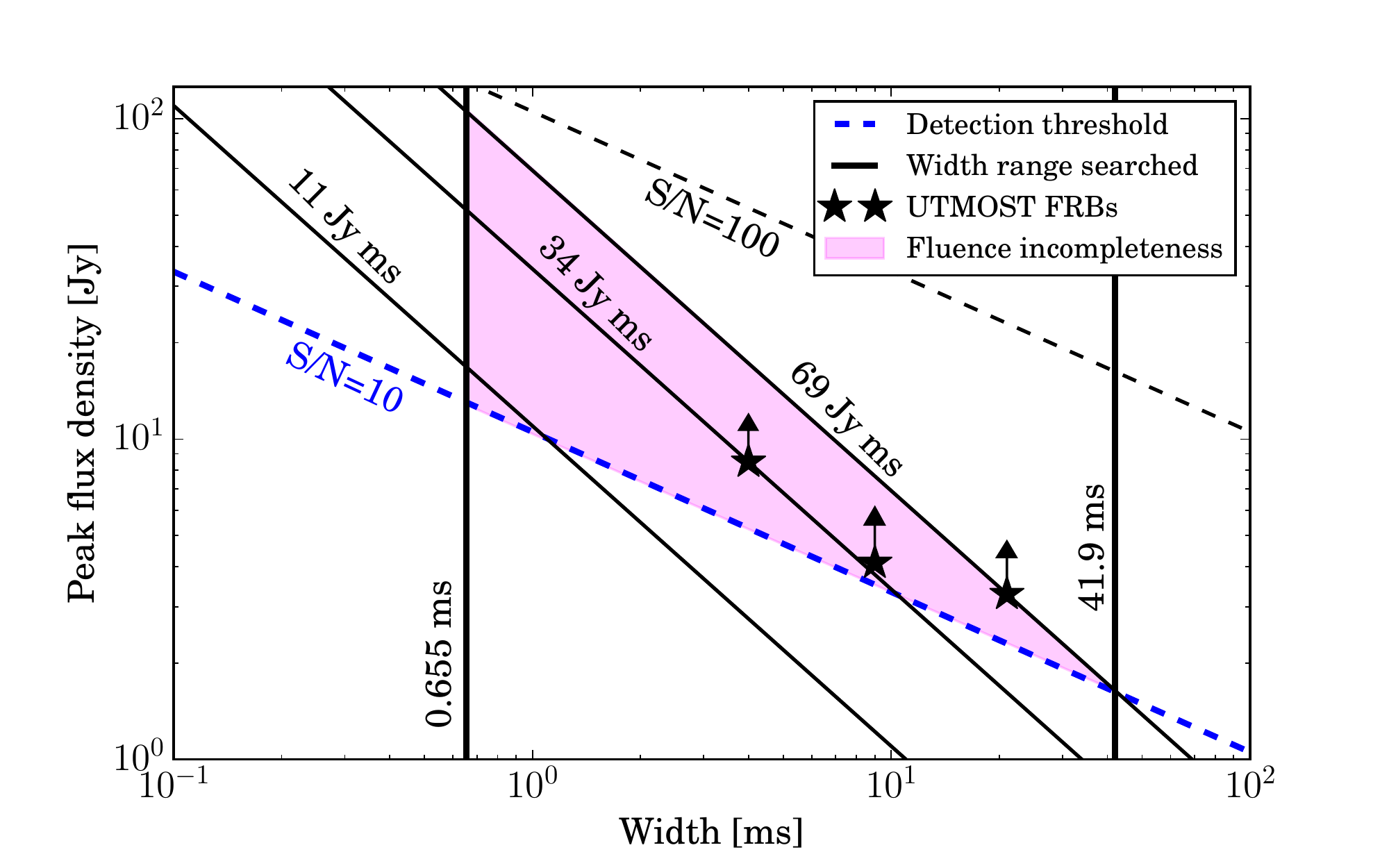}
\caption{Boresight corrected peak flux density versus observed width for the 3 FRBs. Dashed lines
represent lines of constant S/N and solid lines represent lines of constant
fluence. The range of widths searched is enclosed by the solid vertical lines. 
The pink region is the fluence incomplete region which indicates that pulses
with the same fluence but different widths are not equally detectable. 
Only pulses above 69 Jy ms are detectable across the entire width range searched
at UTMOST.}
\label{fig:fluence}
\end{figure*}

\begin{figure*}
\includegraphics[width=5.0 in]{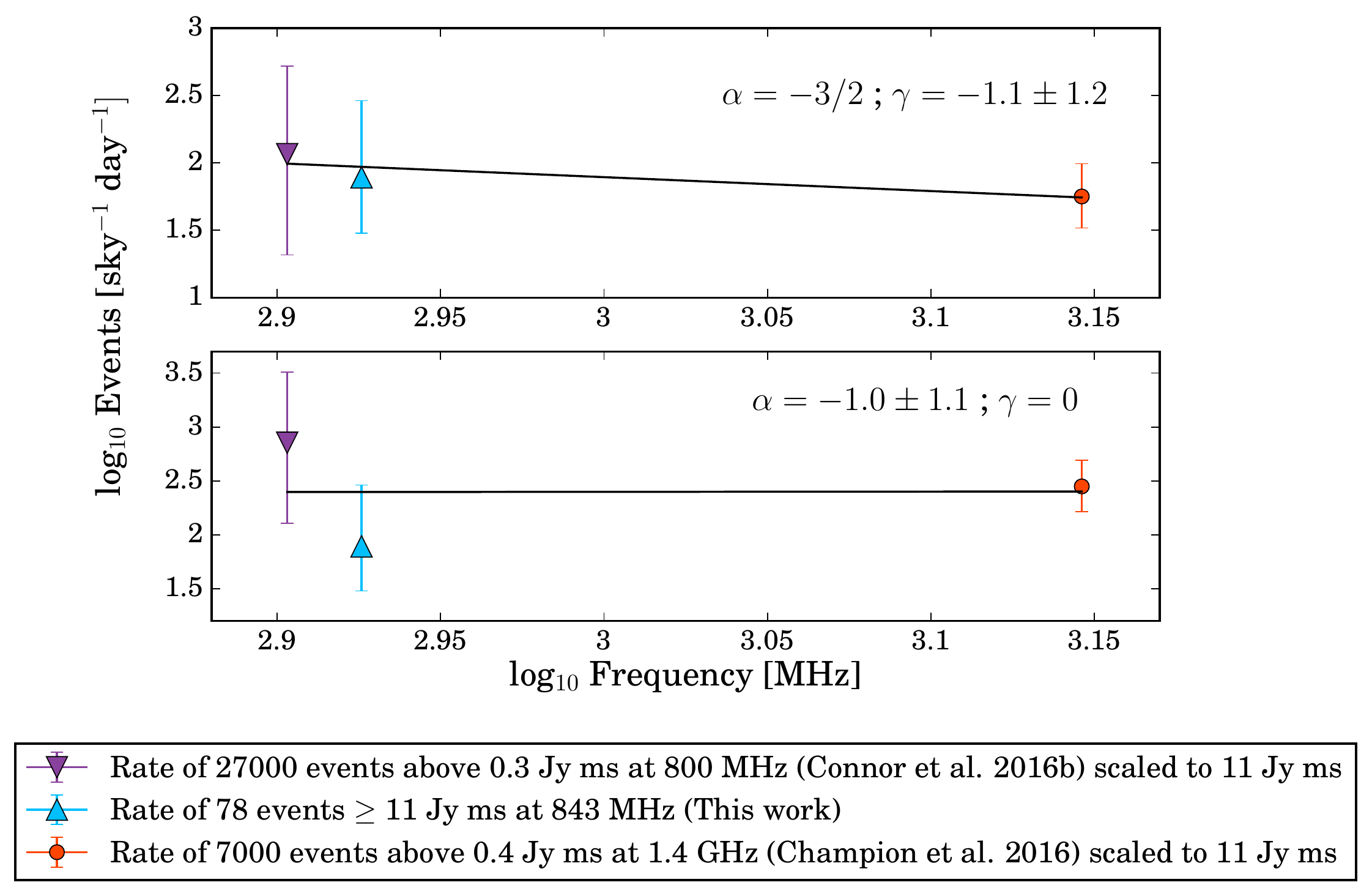}
\caption{All-sky rates at the GBT and Parkes telescope, scaled to the measured UTMOST 
fluence limit of 11 Jy ms. For events 
to be detectable at UTMOST's present
sensitivity, either 1) the spectral index should be steep or 2) the slope
of log$N$-log$\mathcal{F}$ should be flat. Top panel: Constraint on 
$\gamma$ assuming $\alpha=-3/2$. 
Bottom panel: Best fit for $\alpha$ assuming $\gamma=0$.}
\label{fig:spectral_index}
\end{figure*}

\subsection{Localisation}
All 3 FRBs were discovered while following up known
sources, which were at ``bore-sight'', and thus centred on Beam
  177. We localise each FRB's position on the sky, using the angular
  separation between the FRB's position in fan-beam space from the
  bore-sight fan-beam. 
  The 1$\sigma$ uncertainty in the
  direction of the semi-major axis is defined by the primary beam
  ($\sim 2.8^{\circ}$) while the uncertainty in the semi-minor axis is
  controlled by the fractional S/N measured from the adjacent fan-beam
  detections.
  For FRBs detected in adjacent fan-beams,
  we perform a simple linear interpolation based on the S/N to localise the event in the fan-beam grid. For the FRB
  with a single fan-beam detection only, we assume the centre of the fan-beam for the localisation. This allows us to construct a
  trace on the sky relative to the bore-sight
  (RAJ, DECJ), taking into account the meridian angle of the observation
  at the time of the FRB, and the known, slight skew and tilt of the
  East-West arms relative to true East-West and horizontal. The trace
  is a strong function of the hour angle of the observation, as one
  would expect in an East-West array. 
The uncertainty in the direction of the semi-minor axis was confirmed observationally
using single pulse detections from bright pulsars. Single pulses that 
were detected in two adjacent fan-beams with S/N similar to our FRBs 
were chosen to estimate our localisation accuracy on the sky. 
These fan-beam localisations were then compared 
to the true position of the pulsar. The 1$\sigma$ scatter of the calculated 
position of the pulsar from individual pulses, compared to the known 
position of the pulsar, is $\sim 0.1$ fan-beams or $\sim 5\arcsec$, in the
direction perpendicular to the fan-beam. For single fan-beam detections at low S/N, 
a similar analysis yielded a slightly poorer localisation precision of 7$\arcsec$.
Two of the FRBs found with UTMOST have FWHP error ellipses of 5$\arcsec \times 2.8^{\circ}$ ($\sim 11$
arcmin$^2$) on the sky as seen, in Figure \ref{fig:frbspos}.
For comparison, single beam FRB detections (with FWHP beamwidth of 14.0$\arcsec$) at 
Parkes, are localised to $\sim160$ arcmin$^2$.
The probability density of the
localisation is shown in right ascension (RA) and declination (DEC),
with the cross marking the most likely position of the burst for each
FRB. 

One of the advantages of UTMOST is that we can localise pulses
to a few arcsec radius if the source is observed at different hour
angles. The tilt of the error ellipses seen in Figure
\ref{fig:frbspos} demonstrates this and is a result of the geometry
of the telescope. The most likely position of the FRB is marked by the cross.
The fan-beams rotate in position according to the
hour angle pointing of the telescope. A repeat FRB pulse, at a
significantly different hour angle, 
allows us to localise the bursts to a few
arcsec, depending on the S/N of the event. We have tested
this scenario using individual pulses from four different
pulsars, with results for four different hour angles (and
offsets from boresight) for the Vela pulsar shown in Figure \ref{fig:vela}.
Localisations to about 5${\arcsec}$ accuracy, or 0.1 fan-beam widths,
should be possible for a bright, repeating FRB. UTMOST is part of the
shadowing campaign of the Survey for Pulsars and Extragalactic Radio
Bursts (SUPERB) at the Parkes radio telescope (Keane et al. in prep). 
An FRB detected simultaneously with Parkes and UTMOST will yield a 
good localisation even if the burst does not repeat.
The sky positions of the 3 FRBs were also re-observed at different
HAs, to look for additional bursts. We spent 105 hours following
FRB 160317, 43 hours on FRB 160410 and 35 hours on FRB 160608. 
The data were searched offline for pulses with S/N $\geq10$ and
with $\pm20\%$ of the DM of the FRB, using the pipeline described in
Section \ref{sec:pipeline}. No repeat pulses were found from any of
the FRB positions.

\section{FRB event rate at UTMOST}
\label{sec:disc}

\subsection{Event rate analysis}
\label{sec:rate}

\cite{Liam} present detailed analyses constraining FRB rates at various telescopes,
scaling from a single FRB discovered at 800 MHz at the GBT (FRB 110523) \citep{Masui}. They estimate
a rate of $4.2^{+19.6}_{-3.2} \times 10^{-1}$ d$^{-1}$ at UTMOST operating at its design sensitivity, based on comparisons between the sensitivities 
and areas surveyed by the two telescopes. This estimate is consistent with \cite{Caleb}.
Using the same method as outlined in \cite{Liam} we calculate a rate of $0.014^{+0.05}_{-0.013}$ d$^{-1}$, which agrees with the
rate we measure at UTMOST.
We have spent a total of 180 days on sky and discovered 3 FRBs with a FoV of
8.8 deg$^{2}$. 
Based on this, we measure a detectable event rate of ($\mathcal{R}$),

\begin{equation}
\label{eq:rate}
\mathcal{R}\, (\mathcal{F} \gtrsim 11 \, \mathrm{Jy\, ms}) \gtrsim 0.78^{+1.24}_{-0.57} \, \times 10^{2} \, \mathrm{events \,\, sky^{-1} \, d^{-1}}
\end{equation}

\noindent at the 95 percent confidence level \citep{Gehrels}, above a full power boresight fluence of 11 Jy ms as parametrised by Equation \ref{eq:lim_flue},
at the half-power FoV.
The rate is given as a lower limit since all searches are
incomplete in the fluence-width plane.
Following \cite{KeanePetroff}, our fluence complete rate is

\begin{equation}
\mathcal{R}\, (\mathcal{F} \gtrsim 69  \, \mathrm{Jy\, ms}) \sim 5.0^{+18.7}_{-4.7}  \, \mathrm{events \,\, sky^{-1} \, d^{-1}},
\end{equation}

\noindent as shown in Figure \ref{fig:fluence}.

In \cite{Caleb_sim,Caleb} we have made estimates of the event
rate expected at UTMOST, scaling from the event rate at Parkes, under
assumptions about the integral source count distribution
(log$N$-log$\mathcal{F}$ relation) and the spectral index of
FRBs. To do this we assume that the spectral energy distribution is flat
between the Parkes 1.4 GHz and UTMOST's 843 MHz operating frequencies,
and that the source count distribution scales as,

\begin{equation}
N(>\mathcal{F}_\mathrm{lim}) = \mathcal{F}_\mathrm{lim}^{\alpha}
\end{equation}

\noindent where $\alpha=-3/2$ for events
populated in a Euclidean Universe.  Under these assumptions,
\cite{Caleb} predict a rate of $\gtrsim0.008(0.004)$ events d$^{-1}$
for a 10$\sigma$, 1-ms wide FRB to a minimum detectable 
fluence of $\mathcal{F}_\mathrm{lim} = 11$ Jy ms at boresight (see Equation \ref{eq:lim_flue}).
If we correct this rate for the measured primary beam dimensions being 10\% larger (Bailes et al., submitted) than adopted in 
\cite{Caleb} we get a rate of $\gtrsim0.007(0.004)$ events d$^{-1}$
which translates to being able to detect 1.3 events in 180 days on sky.
This is in mild tension with our discovery of 3 events in the survey.
We quantify this tension by calculating the probability of observing 3 or more events
to be 14.3\%, assuming Poisson statistics with a mean of 1.3.
Note that the predicted rate at UTMOST takes into account
pulse-width broadening in the current
implementation of the backend (channel widths $\sim780$ kHz) but does not 
account for possibly highly atypical scintillation properties
along specific lines of sight to FRB events. 
The difference in the estimated and measured rates
could be due to FRBs being brighter than expected at 843 MHz,
and/or the slope of the source count distribution $\alpha$, being
shallower than the assumed value. Simultaneous broad-band detection of
an FRB (e.g. Parkes + UTMOST) would help constrain the spectral index and
resolve the question. 
In \cite{Caleb_sim}, we measured $\alpha \approx -0.9 \pm 0.3$, from 9
FRBs discovered in the high latitude sub-survey of the HTRU survey at Parkes. This is consistent with the
events occurring at cosmological distances in a $\Lambda$CDM Universe,
in which the Euclidean value for $\alpha$ does not hold.
Assuming a flat spectral index for FRBs ($\gamma=0$), if we scale the rate at UTMOST from Equation 
\ref{eq:rate}, assuming $\alpha=-1.0$ for the slope of the log$N$-log$\cal{F}$ relation based on the best-fit from the
bottom panel in Figure \ref{fig:spectral_index}, we obtain 
a rate of $\gtrsim 2.1 \times 10^{3}$ events sky$^{-1}$ d$^{-1}$ at 1.4 GHz, consistent with the observed Parkes
rate at the 2$\sigma$ level \citep{Champion}.

{\subsection{Constraints on spectral and source count distribution indices}

As discussed in the previous section, the observed FRB rate at UTMOST
at 843 MHz can be brought to consistency with the rate found at 1.4 GHz at Parkes if
we assume FRBs are flat spectrum sources on average, and that the
log$N$-log$\mathcal{F}$ relation has slope $\alpha = -1.0$ 
(best-fit for $\alpha$ assuming $\gamma = 0$, in bottom panel of Figure \ref{fig:spectral_index}).
Alternatively, we can relax the flat spectrum assumption, and ascribe
the higher than expected rate to FRBs being brighter at 843 MHz than
at 1.4 GHz. Assuming the log$N$-log$\mathcal{F}$ relation has slope
$\alpha$ and that the FRBs have a power law spectral energy
distribution with index $\gamma$ such that $S \propto \nu^{\gamma}$,
we examine the following scenarios:

\begin{enumerate}

\item \textit{$\alpha=-3/2$}: Based on the detection rates at the
Parkes, GBT and UTMOST telescopes, we can
constrain a spectral index for FRBs as shown in Figure
\ref{fig:spectral_index}. The rate of $\sim 7000$ events sky$^{-1}$
d$^{-1}$ at 1.4 GHz at Parkes, above a fluence limit of 0.4 Jy ms
\citep{Champion} scales to
$\sim56$ events sky$^{-1}$ d$^{-1}$ above UTMOST's fluence limit of 11 Jy ms, and
assuming the spectral index to be flat ($\gamma=0$) 
(see Section \ref{sec:rate}). Similarly, the rate of $\sim 2.7 \times 10^{4}$
events sky$^{-1}$ d$^{-1}$ at 800 MHz at the GBT, above a fluence
threshold of 0.3 Jy ms \citep{Liam} and scaled to UTMOST's fluence
threshold of 11 Jy ms is $\sim116$ events sky$^{-1}$
d$^{-1}$. Using these values we fit for the FRB spectra to be
$\gamma = -1.1(1.2)$ (top panel of Figure \ref{fig:spectral_index}).
This is found to be consistent with the observed constraint in 
\cite{nat_keane} albeit for just one FRB.

\item \textit{$\gamma=0$}: A similar analysis can be done to
constrain the index $\alpha$ of the integral source count distribution assuming
a flat spectral index. We constrain a value of $\alpha=-1.0(1.1)$ for 
$\gamma=0$ (bottom panel of Figure
\ref{fig:spectral_index}). This value of $\alpha$ gives scaled rates of $\sim 270$ events 
sky$^{-1}$ d$^{-1}$ at 1.4 GHz 
at Parkes and $\sim 690$ events sky$^{-1}$ d$^{-1}$ at 800 MHz at 
GBT.

\end{enumerate}

Another possible scenario is that FRBs are giant 
pulses from  pulsars \citep{Wasserman}. The average observed spectral index for 
pulsars is $\gamma = -1.6$ (Jankowski et al., in submitted). If we assume this to be typical of FRBs, 
we fit a slope of $\alpha = -1.76$ for their log$N$-log$\cal{F}$ distribution.
We note that the repeat FRB pulses from the
Arecibo FRB 121102 exhibit a wide range of spectral indices 
\citep[$\gamma \sim \, -10 \, \mathrm{to} \,+14$ ;][]{nat_spitler}, 
similar to giant pulses from the Crab pulsar. For example giant pulses 
from the Crab pulsar exhibit spectral volatility in their broad range 
of spectral indices 
\citep[$\gamma \sim  \, -15 \, \mathrm{to} \, +10$ ;][]{Karuppusamy},
therefore it will be difficult to estimate the mean of the spectral indices until 
the numbers are sufficiently high.
From simultaneous observations of FRB 150418 with Parkes at 1.4 GHz 
and the Murchison Widefield Array (MWA) at 150 MHz \citep{nat_keane}, the non-detection 
at the MWA places a limit of $\gamma > -3.0$. \cite{Vfastr} 
estimate a weak constraint of $-7.6 < \gamma < 5.8$ based on the detection 
sensitivity of \cite{Champion} which is consistent with our estimated 
values. These constraints are only valid if the spectral energy 
distribution (SED) is an unbroken broadband power law and insignificantly
affected by scintillation. This remains to be observationally proven. 
Future broadband instruments like CHIME \citep{Bandura} should have high 
FRB discovery rates and spectral coverage to definitively test this.
Using the method in \cite{Caleb} we scale the observed rate 
at UTMOST for a boresight fluence of 11 Jy ms, to estimate the rates at CHIME and HIRAX \citep{Newburgh}
under a Euclidean Universe assumption. We expect CHIME to detect $\sim 70$ events beam$^{-1}$ d$^{-1}$
for $T_\mathrm{sys} = 50$ K, S/N $= 10$, $G = 1.38$ K Jy$^{-1}$, $N_\mathrm{p} = 2$ and FoV = 250 deg$^{2}$ \citep{Connor,Ng}. Similarly, we expect 
350 events beam$^{-1}$ d$^{-1}$ at HIRAX for $T_\mathrm{sys} = 50$ K, S/N $= 10$, $G = 10.5$ K Jy$^{-1}$, $N_\mathrm{p} = 2$
and FoV = 56 deg$^{2}$ \citep{Newburgh}.

\section{Conclusions}
\label{sec:conc}

In this paper we present the first interferometric detections of FRBs, 
found during 180 days on sky at UTMOST. The events are
beyond the $\approx 10^4$ km near-field limit of the telescope, ruling out local
(terrestrial) sources of interference as a possible origin. We
demonstrate with pulsars that a repeating FRB seen at UTMOST has
the potential to be localised to $\approx 15\arcsec$ diameter error
circle, an exciting prospect for identifying the host.

An all-sky rate of $\mathcal{R}(\gtrsim 11$ Jy\,ms) $\gtrsim
0.78^{+1.24}_{-0.57} \times 10^{2}\, \mathrm{events \,\, sky^{-1} \,
d^{-1}}$ at 843 MHz is calculated from our 3 events, at the boresight fluence out to the
half-power FoV.
Based on the time spent on sky and the number of detections made, we
measure a rate of $0.017^{+0.03}_{-0.01}$ events
beam$^{-1}$ d$^{-1}$ at UTMOST, for the sensitivity achieved during the upgrade.
The rates estimated by \cite{Caleb} for
the present sensitivity, is only 14.3\% unlikely assuming Poisson statistics with 
a mean of 1.3. One possibility could be due to the log$N$-log$\cal F$ relation for 
events being flatter than for a Euclidean Universe, having a slope $\alpha \approx -1.0$, 
rather than $\alpha = -3/2$. In this case, searching for FRBs with a less sensitive, but wider field
of view instrument, appears to be a competitive strategy,
\cite[e.g.][]{Vedantham}. Alternatively, FRBs may simply be brighter at
843 MHz on average than at 1.4 GHz, implying a steeper spectral index
for FRBs. Assuming a Euclidean Universe scaling, we
find a best fit spectral index of $\gamma = -1.1(1.2)$.}
Our ongoing work, and the work of others at many other facilities, will 
settle these questions once sufficient numbers of FRBs are detected over a 
broad frequency range. Understanding the spectra and log$N$-log$\mathcal{F}$ 
distributions are vital in the quest to understand this enigmatic population.

\section*{Acknowledgments}
 
The authors would like to thank the referee for their insightful suggestions. We would also like to thank Jr-Wei Tsai and Liam Connor for useful discussions.
The Molonglo Observatory is owned and operated by the University of 
Sydney with support from the School of Physics. 
The upgrade to the observatory has been supported by the University of Sydney through the Federation Fellowship FF0561298 
and the Science Leveraging Fund of the New South Wales Department of Trade and Investment.
Support for operations and continuing development of the observatory is provided by
the University of Sydney, Swinburne University of Technology, the Australian Research Council 
Centre for All-Sky Astrophysics (CAASTRO), through project number CE110001020, and the 
Laureate Fellowship FL150100148. The late Professor George Collins allocated strategic funds for the purchase of the supercomputer in
use at the facility from Swinburne University and was a passionate advocate for the project.


\bibliographystyle{mnras}
\bibliography{UTMOST_FRBs}


\appendix

\section{Positional coordinates of FRBs 160317, 160410 and 160608}
The coordinates of the FRB localisation ellipses in Figure \ref{fig:frbspos} are given in Table \ref{table:coords}.

\begin{table*}
\caption{Sky coordinates of the 3 UTMOST FRBs. For each FRB, the first two columns are the J2000 right ascensions (RAs) and declinations (DECs) 
of the coordinates of the line defining the major axis of the $3\sigma$ FRB localisation contours in Figure 
\ref{fig:frbspos}, given in units of degrees. The third column gives the probability of the event occurring at this point along the localisation arc.} 
\label{table:coords} 
\centering
\begin{tabular}{c c c | c c c | c c c }
\hline\hline
\multicolumn{3}{c}{FRB 160317}  &  \multicolumn{3}{c}{{FRB 160410}} &  \multicolumn{3}{c}{{FRB 160608}} \\ [0.5ex] 
RA (hrs) & DEC (deg) & Prob. & RA (hrs) & DEC (deg) & Prob. & RA (hrs) & DEC (deg) & Prob. \\
\hline \\ [0.5ex]
7.9356 & $-32.4208$ & 0.0027 & 8.6865 & 3.3129 & 0.0028 & 7.6415 & $-43.5684$ & 0.0031\\
7.9332 & $-32.2575$ & 0.0038 & 8.6867 & 3.4761 & 0.0039 & 7.6396 & $-43.4052$ & 0.0043\\
7.9308 & $-32.0943$ & 0.0053 & 8.6869 & 3.6394 & 0.0054 & 7.6378 & $-43.2419$ & 0.0058\\
7.9284 & $-31.9310$ & 0.0072 & 8.6870 & 3.8027 & 0.0072 & 7.6359 & $-43.0786$ & 0.0077\\
7.9260 & $-31.7677$ & 0.0095 & 8.6872 & 3.9659 & 0.0096 & 7.6341 & $-42.9154$ & 0.0101\\
7.9237 & $-31.6045$ & 0.0124 & 8.6874 & 4.1292 & 0.0123 & 7.6323 & $-42.7521$ & 0.0129\\
7.9213 & $-31.4412$ & 0.0159 & 8.6876 & 4.2924 & 0.0156 & 7.6305 & $-42.5888$ & 0.0163\\
7.9190 & $-31.2780$ & 0.0199 & 8.6878 & 4.4557 & 0.0194 & 7.6287 & $-42.4256$ & 0.0200\\
7.9167 & $-31.1147$ & 0.0243 & 8.6881 & 4.6190 & 0.0236 & 7.6269 & $-42.2623$ & 0.0242\\
7.9145 & $-30.9514$ & 0.0291 & 8.6883 & 4.7822 & 0.0282 & 7.6252 & $-42.0991$ & 0.0287\\
7.9122 & $-30.7882$ & 0.0342 & 8.6885 & 4.9455 & 0.0330 & 7.6234 & $-41.9358$ & 0.0334\\
7.9100 & $-30.6249$ & 0.0393 & 8.6888 & 5.1088 & 0.0378 & 7.6217 & $-41.7725$ & 0.0381\\
7.9078 & $-30.4616$ & 0.0442 & 8.6890 & 5.2720 & 0.0425 & 7.6200 & $-41.6093$ & 0.0426\\
7.9056 & $-30.2984$ & 0.0486 & 8.6893 & 5.4353 & 0.0468 & 7.6183 & $-41.4460$ & 0.0467\\
7.9034 & $-30.1351$ & 0.0524 & 8.6895 & 5.5986 & 0.0505 & 7.6166 & $-41.2827$ & 0.0502\\
7.9012 & $-29.9718$ & 0.0554 & 8.6898 & 5.7618 & 0.0535 & 7.6149 & $-41.1195$ & 0.0530\\
7.8991 & $-29.8086$ & 0.0572 & 8.6901 & 5.9251 & 0.0555 & 7.6133 & $-40.9562$ & 0.0548\\
7.8970 & $-29.6453$ & 0.0579 & 8.6904 & 6.0884 & 0.0564 & 7.6116 & $-40.7929$ & 0.0556\\
7.8949 & $-29.4820$ & 0.0574 & 8.6907 & 6.2516 & 0.0562 & 7.6100 & $-40.6297$ & 0.0553\\
7.8928 & $-29.3188$ & 0.0557 & 8.6910 & 6.4149 & 0.0549 & 7.6084 & $-40.4664$ & 0.0540\\
7.8907 & $-29.1555$ & 0.0529 & 8.6913 & 6.5782 & 0.0526 & 7.6068 & $-40.3031$ & 0.0517\\
7.8887 & $-28.9922$ & 0.0492 & 8.6916 & 6.7414 & 0.0493 & 7.6052 & $-40.1399$ & 0.0485\\
7.8866 & $-28.8290$ & 0.0448 & 8.6919 & 6.9047 & 0.0454 & 7.6036 & $-39.9766$ & 0.0447\\
7.8846 & $-28.6657$ & 0.0400 & 8.6923 & 7.0680 & 0.0409 & 7.6021 & $-39.8133$ & 0.0404\\
7.8826 & $-28.5024$ & 0.0349 & 8.6926 & 7.2312 & 0.0362 & 7.6005 & $-39.6501$ & 0.0357\\
7.8806 & $-28.3392$ & 0.0299 & 8.6930 & 7.3945 & 0.0313 & 7.5990 & $-39.4868$ & 0.0310\\
7.8787 & $-28.1759$ & 0.0250 & 8.6933 & 7.5578 & 0.0266 & 7.5975 & $-39.3235$ & 0.0265\\
7.8767 & $-28.0126$ & 0.0205 & 8.6937 & 7.7210 & 0.0221 & 7.5960 & $-39.1603$ & 0.0221\\
7.8748 & $-27.8494$ & 0.0165 & 8.6941 & 7.8843 & 0.0181 & 7.5945 & $-38.9970$ & 0.0181\\
7.8729 & $-27.6861$ & 0.0130 & 8.6945 & 8.0476 & 0.0144 & 7.5930 & $-38.8337$ & 0.0146\\
7.8710 & $-27.5229$ & 0.0100 & 8.6949 & 8.2108 & 0.0113 & 7.5915 & $-38.6705$ & 0.0115\\
7.8691 & $-27.3596$ & 0.0075 & 8.6953 & 8.3741 & 0.0087 & 7.5900 & $-38.5072$ & 0.0089\\
7.8672 & $-27.1963$ & 0.0056 & 8.6957 & 8.5373 & 0.0065 & 7.5886 & $-38.3440$ & 0.0067\\
7.8654 & $-27.0331$ & 0.0040 & 8.6961 & 8.7006 & 0.0048 & 7.5871 & $-38.1807$ & 0.0050\\
7.8635 & $-26.8698$ & 0.0029 & 8.6965 & 8.8639 & 0.0035 & 7.5857 & $-38.0174$ & 0.0037\\

\hline  
\end{tabular} 
\end{table*}

\end{document}